\documentclass[a4paper,11pt]{article}
\usepackage{aaskaiid}
\usepackage{orcidlink}
\usepackage{booktabs}
\newcommand{\kms}{\mbox{km~s$^{-1}$}}
\newcommand{\cmc}{\mbox{cm$^{-3}$}}
\newcommand{\Hi}{\mbox{H\,\textsc{i}}}
\newcommand{\Htwo}{\mbox{H$_2$}}
\newcommand{\Msun}{\mbox{$M_\odot$}}

\title{Low-Frequency Recombination Lines from Galaxies and AGN over Cosmic Time}
\ShortTitle{Extragalactic Low-Frequency RRLs}
\ShortName{Emig et al.} 

\author[1]{Kimberly L.~Emig\orcidlink{0000-0001-6527-6954}}
\affiliation[1]{National Radio Astronomy Observatory, 520 Edgemont Road, Charlottesville, VA 22903, USA}
\emailAdd{kemig@nrao.edu}

\author[2]{Sergei Balashev}
\affiliation[2]{Department of Theoretical Astrophysics, Ioffe Institute, Politechnicheskaya ul. 26, Saint-Petersburg, Russia}

\author[3]{Francoise Combes}
\affiliation[3]{Observatoire de Paris, LUX, Coll\`ege de France, CNRS, PSL University, Sorbonne University, 75014, Paris, France}

\author[4]{Lucie Cros}
\affiliation[4]{Laboratoire de Physique de l’École Normale Supérieure, ENS,
Université PSL, CNRS, Sorbonne University, University Paris Cité, 75005 Paris, France}

\author[1]{Bjorn Emonts}

\author[5]{Neeraj Gupta\orcidlink{0000-0001-7547-4241}}
\affiliation[5]{Inter-University Centre for Astronomy and Astrophysics, 
Post Bag 4, Ganeshkhind, Pune 411 007, India}

\author[4]{Antoine Gusdorf\orcidlink{0000-0002-0354-1684}}

\author[8]{Emmanuel Momjian\orcidlink{0000-0003-3168-5922}}
\affiliation[8]{National Radio Astronomy Observatory, 1011 Lopezville RD., Socorro, NM 87801, USA}

\author[9]{S\'ebastien Muller \orcidlink{0000-0002-9931-1313}}
\affiliation[9]{Department of Space, Earth and Environment, Chalmers University of Technology, Onsala Space Observatory, SE-43992 Onsala, Sweden}

\author[10,11]{Elaine M. Sadler}
\affiliation[10]{Sydney Institute for Astronomy, School of Physics A28, University of Sydney, NSW 2006, Australia} \affiliation[11]{ATNF, CSIRO Space and Astronomy, PO Box 76, Epping, NSW 1710, Australia}

\author[1]{Pedro Salas}

\author[12]{Alexander G. G. M. Tielens}
\affiliation[12]{University of Maryland, Department of Astronomy, Room 1113 PSC Bldg. 415, College Park, MD 20742-2421, USA}

\author[1,13]{Ilsang Yoon}
\affiliation[13]{Department of Astronomy, University of Virginia, P.O. Box 3818, Charlottesville, VA 22903, USA}

\abstract{Radio recombination lines (RRLs) at low frequencies ($\lesssim$10 GHz) can provide a multi-phase view of interstellar gas in nearby galaxies, absorption-line-systems, and AGN. Hydrogen RRLs arise in fully ionized gas and carbon RRLs trace elusive cold-HI and CO-dark molecular gas. Low frequency RRLs are typically stimulated by the radio continuum and thus may be observable within or against radio bright sources out to cosmological distances ($z \sim 6$). Although long sought after, RRLs were only recently detected outside of the local universe \citep[at $z \sim 1$;][]{Emig2020a, Emig2023}. Such detections have been made possible by the advancement of wide-bandwidth spectral-line surveys on next-generation low-frequency telescopes. Precursors and pathfinders to the SKA have opened up this field of research and will make significant advancements over the next years by enabling surveys over  large source samples. The SKA will provide access to the crucial frequency ranges where RRL line intensity is brightest. Furthermore, multi-band SKA measurements will fully characterize gas physical conditions. Key extragalactic science of low frequency RRLs will focus on (i) the conversion of baryonic material into stars across cosmic time, (ii) the evolution of the ISM and its physical conditions in galaxies, and (iii) how gas drives and inhibits AGN activity.}

\begin{document}
\newcommand{\actaa}{Acta Astron.} 
\newcommand{\araa}{ARA\&A} 
\newcommand{\aar}{A\&ARv} 
\newcommand{\aapr}{A\&ARv} 
\newcommand{\ab}{Astrobiol.} 
\newcommand{\aj}{AJ} 
\newcommand{\apj}{ApJ} 
\newcommand{\apjl}{ApJL} 
\newcommand{\apjs}{ApJSS} 
\newcommand{\ao}{Appl. Opt.} 
\newcommand{\apss}{Astro. \& Space Sci.} 
\newcommand{\aap}{A\&A} 
\newcommand{\aaps}{A\&AS.} 
\newcommand{\baas}{Bull. Am. Astron. Soc.} 
\newcommand{\caa}{Chinese A\&A} 
\newcommand{\cjaa}{Chinese J. A\&A} 
\newcommand{\cqg}{Class. Quantum Gravity} 
\newcommand{\gal}{Galaxies} 
\newcommand{\gca}{Geo. Cosmo. Acta} 
\newcommand{\icarus}{Icarus} 
\newcommand{\jcap}{JCAP} 
\newcommand{\jgr}{J. Geophys. Res.} 
\newcommand{\jgrp}{J. Geophys. Res. Planets} 
\newcommand{\jqsrt}{J. Quant. Spectrosc. Radiat. Transf.} 
\newcommand{\memsai}{Mem. SAIt} 
\newcommand{\mnras}{MNRAS} 
\newcommand{\nat}{Nature} 
\newcommand{\nastro}{Nat. Astron.} 
\newcommand{\ncomms}{Nat. Commun.} 
\newcommand{\nphys}{Nat. Phys.} 
\newcommand{\na}{New Astron.} 
\newcommand{\nar}{New Astron. Rev.} 
\newcommand{\physrep}{Phys. Rep.} 
\newcommand{\pra}{Phys. Rev. A} 
\newcommand{\prb}{Phys. Rev. B} 
\newcommand{\prc}{Phys. Rev. C} 
\newcommand{\prd}{Phys. Rev. D} 
\newcommand{\pre}{Phys. Rev. E} 
\newcommand{\prx}{Phys. Rev. X} 
\newcommand{\prl}{Phys. Rev. Let.} 
\newcommand{\psj}{Planet. Sci. J.} 
\newcommand{\planss}{Planet. Space Sci.} 
\newcommand{\pnas}{Proc. Natl Acad. Sci. USA} 
\newcommand{\procspie}{Proc. SPIE} 
\newcommand{\pasa}{PASA} 
\newcommand{\pasj}{PASJ} 
\newcommand{\pasp}{PASP} 
\newcommand{\rmxaa}{RMXAA} 
\newcommand{\sci}{Science} 
\newcommand{\sciadv}{Sci. Adv.} 
\newcommand{\solphys}{Sol. Phys.} 
\newcommand{\sovast}{Soviet Ast.} 
\newcommand{\ssr}{Space Sci. Rev.} 
\newcommand{\uni}{Universe} 

\maketitle

\section{About (Low Frequency) Radio Recombination Line Emission}
\label{sec:rrl_emission}

When electrons recombine to atoms, spectral lines known as radio recombination lines (RRLs) arise from electronic transitions at high principal quantum numbers, $\mathsf{n}$, cascading to lower energy levels. The difference in energy between these high principal quantum levels is small, thus the radiation emitted has low (radio-wavelength) energy. RRLs are typically observed from hydrogen, carbon, and helium (and only in rather rare circumstances, from other elements).

Hydrogen RRLs (HRRLs) originate generally in fully ionized gas. They are typically the brightest RRL lines at frequencies roughly $\gtrsim$300 MHz or so. Helium RRLs (HeRRLs), typically  $\sim 10\%$ of the HRRL intensity of a given $\mathsf{n}$, also arise in ionized gas. Free from extinction by dust, H and He RRLs are vital tools to spectroscopically image the morphology and kinematics of ionized gas and measure the ionizing radiation from young massive stars, i.e., star formation. Since each HeRRL falls with a constant velocity offset with respect to an HRRL of $-$122~\kms{}, these species may be observed simultaneously given sufficient sensitivity.

Carbon RRLs (CRRLs) arise from cold gas, where carbon is singly ionized. Carbon has a lower ionization potential (11.3 eV) than hydrogen (13.6 eV). Therefore, it can remain ionized in regions where hydrogen is in atomic and/or molecular forms.
CRRLs fall with a constant velocity offset with respect to Hydrogen of -149.6~\kms. CRRLs are typically the brightest RRL species at frequencies roughly $\lesssim$300 MHz or so, for which the processes of dielectronic capture and stimulation significantly enhance CRRL emission, and high free-free continuum opacity leads to suppressed HRRL emission \citep[see][and references therein]{Gordon2002}. CRRLs provide exceptional means to observe the layers of clouds where molecular gas is forming and therefore investigate the interstellar medium's (ISM) role in the formation and evolution of galaxies.

At ``high'' radio frequencies, roughly $\gtrsim$ 10 GHz or so for extragalactic targets (though this can vary greatly from source to source), emission from RRLs may be dominated by spontaneous transitions;
the emission either closely follows a Boltzmann distribution, i.e., is in LTE, or has only a smaller correction factor (0.5-1) to be applied.\footnote{The exceptions to high-frequency RRL emission being in LTE arise when gas is low in density and/or when a continuum source behind the emission region has a large flux.} As RRL intensity depends upon the emission measure, thus electron density squared ($\propto n_e^2$), RRLs (at high frequencies) are more readily detectable from higher density gas with thermal continuum emission that is not yet optically thick.

At ``low'' radio frequencies, roughly $\lesssim 10$~GHz as has been observed in extragalactic sources, non-LTE processes affect the level populations (of electrons in atoms), stimulating and enhancing RRL emission. Stimulated emission dominates when the gas is low in density and/or relatively cool for its phase, a bright continuum source falls background to the emission region, and/or when the free-free continuum of the emission region is close to becoming optically thick. 
With a line intensity that is directly proportional to the continuum intensity, stimulated RRLs are observable out to high-redshift towards bright radio-continuum sources, whereas the spontaneous emission at high frequencies is inversely proportional to distance squared. The non-LTE nature of the emission leaves the line intensity dependent upon the physical conditions of the gas; by measuring the RRL spectral line energy distribution (SLED) over about a decade in frequency (see Figures in Section~\ref{sec:ska_impact}), the characteristic density and temperature of the gas are thus determined.
In the Milky Way, the line broadening of low-frequency RRLs provides insightful physics about the gas physical conditions; in extragalactic sources, the lines are more likely to be Doppler broadened.

The transition between one energy level ($ \Delta \mathsf{n} = 1$), referred to as an $\alpha$-transition, is the brightest in a series\footnote{A series is defined as all transitions which reach final state $\mathsf{n}$ following recombination, i.e., they start at an energy level $>\mathsf{n}$.} and are therefore considered to be the ``default'' RRLs referred to in an observation. An example of the notation typically adopted to refer to RRLs, for example for a Hydrogen transition from energy levels $76 \rightarrow 75$, is H75$\alpha$.

In addition to $\alpha$-transitions, higher order RRL transitions are also accessible and provide further insight (as described in Sec.~\ref{sec:state_of_field}). The relative intensity of higher order transitions --- e.g., $\beta$-, $\gamma$-, and $\delta$-transitions with $\Delta \mathsf{n} = 2,3,4$, respectively ---  is given by $I_{\Delta \mathsf{n}} = (\Delta\mathsf{n} \cdot M_{\Delta\mathsf{n}} /  M_{\alpha}) \cdot I_{\alpha}$, or $I_{\Delta \mathsf{n}=2 \,(3,4)} = 0.28 \,(0.13, 0.07) \cdot I_{\alpha}$
for oscillator strengths $M_{\Delta \mathsf{n} =1,2,3,4} = 0.1908, 0.02633, 0.008106, 0.003492$ \citep{Menzel1968}.

\begin{figure}
    \centering
    \includegraphics[width=0.7\linewidth]{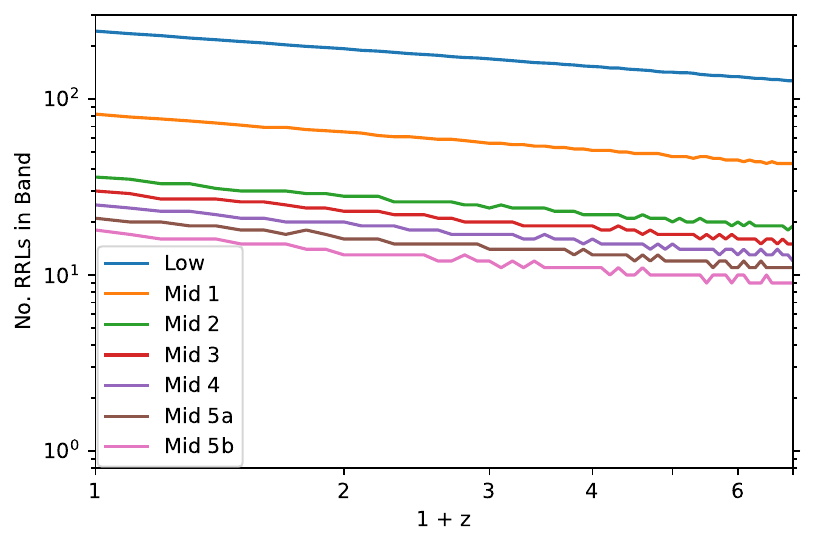}
    \caption{The number of (H$\mathsf{n}\alpha$) RRLs in each SKA band as a function of (1+)redshift, out to $z=6$. }
    \label{fig:rrl_bands_z}
\end{figure}

Table~\ref{tab:ska_cov} shows the (Hydrogen) $\alpha$-, $\beta$-, and $\gamma$-transition RRLs covered in SKA1 Low and Mid Bands at rest frequencies, i.e., a $z=0$ source. It includes 1674 lines, with 1473 accessible to SKA AA*. Also considering the lines from Helium and Carbon, which are effectively simultaneously covered in a spectral setup, the SKA has the incredible (and realistic) potential to detect more than 5000 RRLs. In Figure~\ref{fig:rrl_bands_z}, we show the number of Hydrogen $\alpha$-transitions in each SKA band as a function of redshift.

For an overview of the physics of RRLs and some historical context, we refer the reader to \cite{Gordon2002}.

\begin{table}[]
  \centering
  \caption{SKA coverages of RRLs at rest ($z=0$). \label{tab:ska_cov}}
  \begin{tabular}{lccccccccc}
  \toprule
  Band &Frequency &Chan.~Res. &$\mathsf{n}{\alpha}$ &$N_{\alpha}$ &$\mathsf{n}{\beta}$ &$N_{\beta}$ &$\mathsf{n}{\gamma}$ &$N_{\gamma}$\\
    &(MHz)     &(\kms{}) \\
  \midrule
  Low       &50--350    &32.4--4.6 &508$\alpha$--266$\alpha$   &243    &639$\beta$--334$\beta$ &306    &731$\gamma$--382$\gamma$   &350\\
  Mid 1     &350--1050  &11.5--3.8 &265$\alpha$--184$\alpha$   &82     &333$\beta$--232$\beta$ &102    &381$\gamma$--265$\gamma$   &117\\
  Mid 2     &950--1760  &4.2--2.3  &190$\alpha$--155$\alpha$   &36     &239$\beta$--195$\beta$ &45     &273$\gamma$--223$\gamma$   &51\\
  Mid 3     &1650--3050 &          &158$\alpha$--129$\alpha$   &30     &198$\beta$--162$\beta$ &36     &227$\gamma$--185$\gamma$   &43\\
  Mid 4     &2800--5180 &          &132$\alpha$--108$\alpha$   &25     &166$\beta$--136$\beta$ &31     &190$\gamma$--155$\gamma$   &36\\
  Mid 5a    &4600--8500 &3.5--1.9  &112$\alpha$--92$\alpha$    &21     &140$\beta$--115$\beta$ &26     &160$\gamma$--131$\gamma$   &30 \\
  Mid 5b    &8300--15400&2.9--1.6  &92$\alpha$--75$\alpha$     &18     &115$\beta$--94$\beta$  &22     &131$\gamma$--108$\gamma$   &24 \\
  \bottomrule
  \end{tabular}
   \raggedright 
   Notes: $\mathsf{n}$ denotes the final principal quantum number of a transition.\\
   $\alpha$ ($\beta$, $\gamma$) transitions are those between one (two, three)  energy level(s),  $\Delta \mathsf{n} = 1$~(2, 3). \\
   $N$ refers to the number of respective RRL transitions in the band. \\
   The table has been computed with respect to the rest frequencies of Hydrogen. All transitions of Carbon (Helium) are offset from Hydrogen by -150 (-122) \kms{}. \\
   Some safe general assumptions about the line intensities in a given band: $I_{He} \approx 0.1 I_H$; $I_{\beta} \approx 0.3 I_{\alpha}$; and $I_{\gamma} \approx 0.1 I_{\alpha}$.\\
   The full bandwidth of the `Frequency' range is available at the given `Chan.~Res.' in each `Band' except for SKA-Mid Band 5b which has 5000 MHz of instantaneous bandwidth.
\end{table}

\section{State of the Field: Extragalactic RRL Observations (at Low-Frequencies)}
\label{sec:state_of_field}

Extragalactic RRLs were first detected in the galaxies M82 \citep{Shaver1977, Chaisson1977, Bell1977} and NGC 253 \citep{Seaquist1977}. Soon after, searches for stimulated RRLs towards bright AGN were made -- using single dish telescopes and covering one spectral line per tuning. Three AGN were targeted at 430 MHz to probe CRRLs \citep{Churchwell1979}. An additional 21 (10) AGN were searched for HRRLs at 1.4 (4.8) GHz \citep{Churchwell1979, Bell1980}. The Seyfert 2, GHz-peaked-spectrum galaxy MRK 668 (OQ 208; $D \sim 340$~Mpc) was detected in H99$\alpha$ and H83$\alpha$ using Effelsberg at 6.2 and 10.5 GHz, with line to continuum ratios of $T_L / T_C \approx 3 \times 10^{-3}$ and $\approx 6 \times 10^{-3}$, respectively \citep{Bell1980}; 20 hours on source were spent in each band to obtain these detections. These early searches showed that stimulation at low frequencies was not resulting in exceedingly bright line emission, extragalactic RRL observations required long integrations, and extragalactic RRLs were not exceedingly feasible with available technologies. 

Currently, a total of 24 external galaxies have detections of RRLs\footnote{For an up-to-date list of extragalactic RRL detections, we refer the reader to \href{https://tinyurl.com/DatabaseForExtragalacticRRLs}{tinyurl.com/DatabaseForExtragalacticRRLs} }.  Despite observations overwhelmingly using 5~GHz frequencies or above (and targeting very nearby galaxies), 9 of the 24 sources show evidence for enhanced, non-LTE emission in their RRL spectral line energy distributions (SLEDs). Wide bandwidth receivers (covering many lines in a single observation) on sensitive, high-resolution low-frequency telescopes are making the study of extragalactic RRLs feasible. RRL intensities are well-correlated and vary smoothly across an observing band (see example SLED, Fig.~\ref{fig:DIG_HRRLs}). It is common to stack 20-30 RRLs with instantaneous frequency coverage \citep{Balser2006, Emig2020a}, leading to an increased signal-to-noise by a factor of $\sim$5 through line-stacking. Since the low-frequency-RRL intensity is directly proportional to the continuum intensity, observational sensitivity wins out when targeting the brightest continuum sources.  We describe in more detail select detections that have been reported at low frequencies towards non-local targets.

Using LOFAR 134 MHz observations, \cite{Emig2019} established that it is possible to observe RRLs outside of the local universe, by detecting 13-stacked RRLs at $z=1.1$. 
They targeted 3C 190 ($z_{\mathrm{systemic}}=1.1946$), a radio-bright ($S_{134\,\mathrm{MHz}} \approx 19$~Jy) quasar \citep{Kellermann1964}. As the brightest central galaxy (BCG) of a proto-cluster-like environment, the host galaxy of 3C 190 is undergoing several major and minor mergers \citep{Stockton2001}. Two bright radio jets extend $\sim$22~kpc across, falling just outside of the host galaxy, but one of the jets is interacting with a major merging galaxy. \cite{Ishwara-Chandra2003} detected \Hi{} 21-cm absorption toward an unresolved source, using a GMRT synthesized beam of $\sim5''$. The \Hi{} absorption spans $\sim$600~\kms{} reaching optical depths of $>0.01$. Using stacking  \cite{Emig2019} detected a RRL feature at a redshift of $z=1.124$, corresponding to a 10\,000~\kms{} blue shifted velocity with respect to the systemic redshift of 3C~190. The line width of the feature, FWHM=$31.2 \pm 8.3$~\kms{}, is more typical of cold gas, and the detection likely originates from CRRLs. This work demonstrated that ``high'' redshift studies using RRLs are feasible. To identify the feature as statistically significant a cross-correlation to stack and search for RRLs across redshift space was developed and presented in \cite{Emig2020a}.

Capabilities of HRRL measurements of ionized gas at cosmological distances were then demonstrated with detections at 0.8 and 1.2 GHz towards the radio blazar PKS 1830-211 ($z=2.5$) originating in an intervening galaxy at $z=0.89$ \citep{Emig2023}.
PKS\,1830$-$211 is a radio-bright ($S_{1.4\,\mathrm{GHz}} = 11$\,Jy) blazar ($z=2.5$), whose light is strongly lensed by a $z=0.89$ galaxy. The $z=0.89$ galaxy is an absorption system against PKS\,1830$-$211's continuum and has the most molecular species detected in absorption \citep{Wiklind1996,Muller2011,Muller2014,Combes2021}. This made PKS\,1830$-$211 an obvious choice to use for science and technical verification observations for the MeerKAT Absorption Line Survey \citep[MALS;][]{Gupta2016}.
Using MALS data, \cite{Emig2023} detected RRLs in both the MeerKAT L-band data (865-1712 MHz) by stacking 18 RRLs and in the UHF-band data (544-1088 MHz) by stacking 27 RRLs. The HRRL emission spans $\sim$200~\kms{} and contains two velocity components associated with two lines of sight through the lensing galaxy -- this coincides well with \Hi{} 21 cm and OH 18 cm absorption components. With two measurements of RRL intensities, the density of the gas was constrained to be $n_e = 10^{2.7 \pm 0.6}$\,\cmc{}, from the emission measure of $\log(\frac{EM}{\mathrm{pc\,cm}^{-6}} ) = 4.9^{+0.4}_{-1.7}$; see Section~\ref{sec:ska_impact} for more details on how the RRL observations constrain physical conditions of the gas. These values estimate star formation rate surface densities of $\Sigma_{\mathrm{SFR}} = 10^{-0.2 \pm 0.8}$\,\Msun{}\,yr$^{-1}$\,kpc$^{-2}$. Estimating the cold gas mass surface densities from the MALS detections of \Hi{}\,21\,cm absorption and OH\,18\,cm absorption, showed that this $z=0.89$ galaxy indeed falls on the Kennicutt-Schmidt relation, a relation between the star-formation rate surface density and the cold gas mass surface density.

\subsection{Over the next years}
\label{ssec:next_years}

Over the next years, large-area wide-band spectral-line surveys (i.e., primarily targeting \Hi{} 21cm absorption),  will povide the first insights into the type of objects and conditions which can be probed with HRRLs. The MeerKAT Absorption Line Survey \cite[MALS;][]{Gupta2016} and the First Large Absorption Survey in HI \citep[FLASH;][]{Allison2022} will enable large population searches, investigating thousands of radio sources and their lines-of-sight for the first time (see Table~\ref{tab:rrl_nosources}).
MALS uses MeerKAT's 580--1670 MHz coverage at $\sim$6~\kms{} channel resolution to target roughly 500 pointings centered on compact, bright ($S_{1\,\mathrm{GHz}} > 250$~mJy) AGN in order to investigate \Hi{} and OH absorption out to $z\leq 1.44$ and 2 respectively at deep \citep[1.25\,mJy/b;][]{Deka2024} column density sensitivity ($N_{\mathrm{HI}}$ of a few $10^{19}$\,cm$^{-2}$). FLASH observes with the Australian SKA Pathfinder at 712--1000 MHz and $\sim$6~\kms{} channel resolution; it covers a huge area (24,000 deg$^2$) at comparatively shallower sensitivity \citep[5.5\,mJy/b;][]{Yoon2025} to investigate \Hi{} absorption at $z=0.4-1$ with larger number statistics. Both MALS and FLASH are still currently taking data.

In Table~\ref{tab:rrl_nosources}, we estimate the number of sources that MALS and FLASH can investigate for RRLs upon completion of the surveys. We show the number of (H$\mathsf{n}\alpha$) RRLs covered by each survey at $z=0$ and an approximate sensitivity (using observed RMS noises) after stacking 70\% of the lines, considering a 20 \kms{} channel resolution. We then take the NVSS source counts \citep{Condon1998, Matthews2021}, and integrate for the number of sources which would have a peak $3\sigma$ sensitivity for a line-to-continuum ratio ($S_L / S_C$; i.e., negative optical depth) of $S_L / S_C \leq 10^{-3}$ in a 20~\kms{} channel. We have not taken into account the spatial resolution of each survey, nor have we tried to fold in a redshift distribution for the source population in any way. We note that Milky Way observations and the extragalactic RRL detections thus far detect RRLs with $S_L / S_C \approx 10^{-3} - 10^{-4}$, and they indicate that RRL emission from hydrogen (as opposed to carbon) would be the most prominent at the systems' rest frequencies covered in these surveys.

Table~\ref{tab:rrl_nosources} shows that the number of sources to be investigated are 860 (780) in MALS (FLASH), together more than two orders of magnitude greater than previously searched (10s of sources). This population will investigate intervening, associated, and possibly Galactic RRL emission. Since the frequency spacing between RRLs are unique, searches across redshift can also be conducted \citep[e.g., following methods demonstrated in ][]{Emig2020a}. Detections toward single sources will also guide stacking in groups of sources.

\section{SKA's Impact}
\label{sec:ska_impact}

In the ramp up to the SKA over the next years, results from large-area wide-band spectral-line surveys (see Table~\ref{tab:rrl_nosources}) will gather the first basic information of extragalactic RRLs such as the detection rates and key science that can be done. The SKA will be able to push this into a new frontier, by increasing the number of sources investigated from 1000s to 100\,000s. By going a factor of 4-60 deeper in RRL sensitivity, combined with the increased spatial resolution, it will likely enable new types of phenomena that can be investigated. The SKA is poised to build off of the MALS and FLASH surveys thanks to overlapping sky coverage and its complementary frequency coverage where RRL intensities are expected to be brightest.
While the brightest (i.e., highest flux density) radio AGN are generally at $z \sim 1-2$ \citep[e.g.,][]{Spinrad1985}, SKA can push RRL detections to higher redshift ($z \sim 6$). SquERRLS (Square kilometer array Extragalactic RRL Surveys; see Section~\ref{sec:obs}) will be exceptional survey tools and pave the way for high-resolution and multi-frequency follow up. SKA observations have highest impact potential for CRRL observations of the cool diffuse gas which shine brightest at rest frequencies $<$300 MHz. The channel resolution available at the maximum accessible bandwidth in SKA Low and Mid is primed for extragalactic RRL studies, as no spectral zoom modes are needed to conduct comprehensive surveying.

\begin{table}[htbp]
  \small
  \centering
  \caption{Impacts of (Serendipitous) RRL coverage in current and SKA-proposed wide-band spectral-line surveys \label{tab:rrl_nosources}}
    \begin{tabular}{lcccccccc}
    \toprule
&Frequency &No.   &$\mathsf{n}$ &Spectral       &RRL-stacked    &Spatial &Sky         &No.~Srcs \\
&Coverage  &RRLs  &             &Noise          &Noise          &Resol.  &Cov.        & \\ 
&(MHz)     &(z=0) &(z=0)        &($\mu$Jy b$^{-1}$) &($\mu$Jy b$^{-1}$) &('')    &(deg$^{2}$) &  \\ 
    \midrule
    MALS    & 580--1670 & 67 & 158--224     & 690   & 100    & 14    & 1\,700    & 860 \\
    FLASH   & 712--1000 & 23 & 187--209     & 3000  & 750   & 20    & 24\,000   & 550 \\
    SKA Mid Band 1 & 350--1050 & 82 & 184--265   & 270   & 36    & 2     & 34\,000   & 103\,000 \\
    SKA Low  & 50--350 & 243 & 266--508    & 670   & 51    & 13    & 34\,000   & 168\,000 \\
    \bottomrule
    \multicolumn{8}{l}{\scriptsize ``$\mathsf{n}$'' is the principal quantum numbers of the $\alpha$ ($\Delta \mathsf{n} = 1$) transitions covered over the bandpass for a z=0 source.} \\
    \multicolumn{8}{l}{\scriptsize ``Spectral Noise'' is the rms per 20 \kms{} channel.} \\
    \multicolumn{9}{l}{\scriptsize ``RRL-stacked Noise'' is the rms per 20 \kms{} channel assuming 70\% of the "No.~RRLs" (at z=0) are available to stack.} \\
    \multicolumn{9}{l}{\scriptsize ``No.~Srcs'' is the number of sources that enable a $\geq 3 \sigma$ detection with a line to continuum ratio $\leq 10^{-3}$ within the "Sky Cov."} \\
    \end{tabular}
\end{table}

In the next subsections we emphasize three key areas that SKA observations of RRLs over cosmological time can impact:\\
$\S$~\ref{ssec:HI-to-H2} the \Hi{}-to-\Htwo{} transition across cosmic time,\\
$\S$~\ref{ssec:DIG} metallicity-free measurements of the physical properties of (diffuse) ionized gas,\\
$\S$~\ref{ssec:AGN} the environments and evolution of (compact) AGN.

\subsection{The \Hi{}-to-\Htwo{} transition across cosmic time}
\label{ssec:HI-to-H2}

How do molecular clouds form and evolve? What are the (physical) conditions of molecular gas that is forming or being destroyed -- how might they change over cosmic time?  The gas phases involved in the \Hi{}-to-\Htwo{} transition, cold \Hi{} and CO-dark molecular, are challenging to observe and physically characterize. However, the formation and evolution of molecular gas is vital to the ISM, star formation, and overall evolution of galaxies. Low-frequency CRRLs are a sign post of the \Hi{}-to-\Htwo{} transition; they arise exclusively from cold C$^{+}$ \Htwo{} and/or cold C$^{+}$ \Hi{} layers where molecular gas is forming or being destroyed. 

Observations of low-frequency CRRLs in our Galaxy have shown \citep[for more details see the SKA chapter on Low-frequency RRLs in the Galaxy;][]{Salas01.2026.SKA}, with emission in just one observing band that (1) low-frequency CRRLs (observed in large beams) are coincident with cold \Hi{} through 21~cm \Hi{} self-absorption \citep{Roshi2011, Erickson1995, Kantharia1998a}, and (2) low-frequency CRRL intensities correlate with FUV field \citep{Emig2025}. Furthermore, with multi-band measurements of the line intensities, particularly towards the Cassiopeia A (Cas A) supernova remnant, the gas physical conditions (temperature, density, pathlength) can be determined to within 15\% \citep{Oonk2017, Salas2017, Salas2018}. The morphology and velocity information in comparison to other gas tracers and the physical conditions estimated together with PDR models matching multi-wavelength tracers indicate that the gas towards Cas A is originating from a CO-dark molecular layer \citep{Salas2017, Salas2018, Oonk2017, Chowdhury2019, Cros2025}. Radiative transfer and non-LTE modeling also indicate that the CRRLs should arise from cold ($\propto T^{-2.5}$) and dense ($\propto n^2$) ionized carbon layers of clouds \citep{Salgado2017a, Salgado2017b}. \cite{Salas2019} also showed that just two frequency measurements can constrain the gas pressure well.

Given that many transitions can be observed simultaneously and that their strength depends strongly on the physical conditions, low-frequency CRRLs have huge discovery potential to bring to light the \Hi{}-to-\Htwo{} transition, its physical properties, and how they are evolving over cosmic time. Assessing the density, pressure, and turbulence will bring profound insight into the regulation of star formation, and the evolution of galaxies. Given the many CRRLs present across the band, specific frequencies lost to RFI do not preclude detecting CRRLs at a given redshift.

Other possibilities and synergies include investigating AGN tori through CRRLs; coincident \Hi{} 21-cm absorption \citep{Mahony01.2026.SKA} and precise determinations of column densities from CRRL temperature estimates; synergy with [CII] 158~$\mu$m observations as the cold portion of [CII] emission arises from same CRRLs gas; in nearby galaxies, cool diffuse gas decomposition with HI 21-cm emission, CO and (cold) dust emission studies; and with molecular absorption studies at higher frequencies (e.g., SKA, ALMA2030, ngVLA) or even molecular hydrogen absorptions in the UV.

\subsection{Metallicity-free measurements of the physical properties of (diffuse) ionized gas}
\label{ssec:DIG}

Diffuse ionized gas is a massive component of the ISM, known as the warm ionized medium, which provides an important source of weight to regulate disk galaxies. It reflects stellar and AGN feedback; for example, the thermal pressure of ionized gas helps regulate star formation. Additionally, it may be accreting onto galaxies.

Strengths of low frequency HRRL observations, include: dust-unobscured radio tracer; metallicity independent (as it arises from hydrogen); kinematic tracer from line properties; the physical conditions of the gas can be determined, in particular the density and line-of-sight-integrated pathlength can be disentangled from the emission measure ($EM = n_e^2 \ell $); and, it can provide a means to separate and identify diffuse gas (contaminating) the dense ionized gas directly associated with star formation.

\begin{figure}
    \centering
    \includegraphics[width=0.48\linewidth]{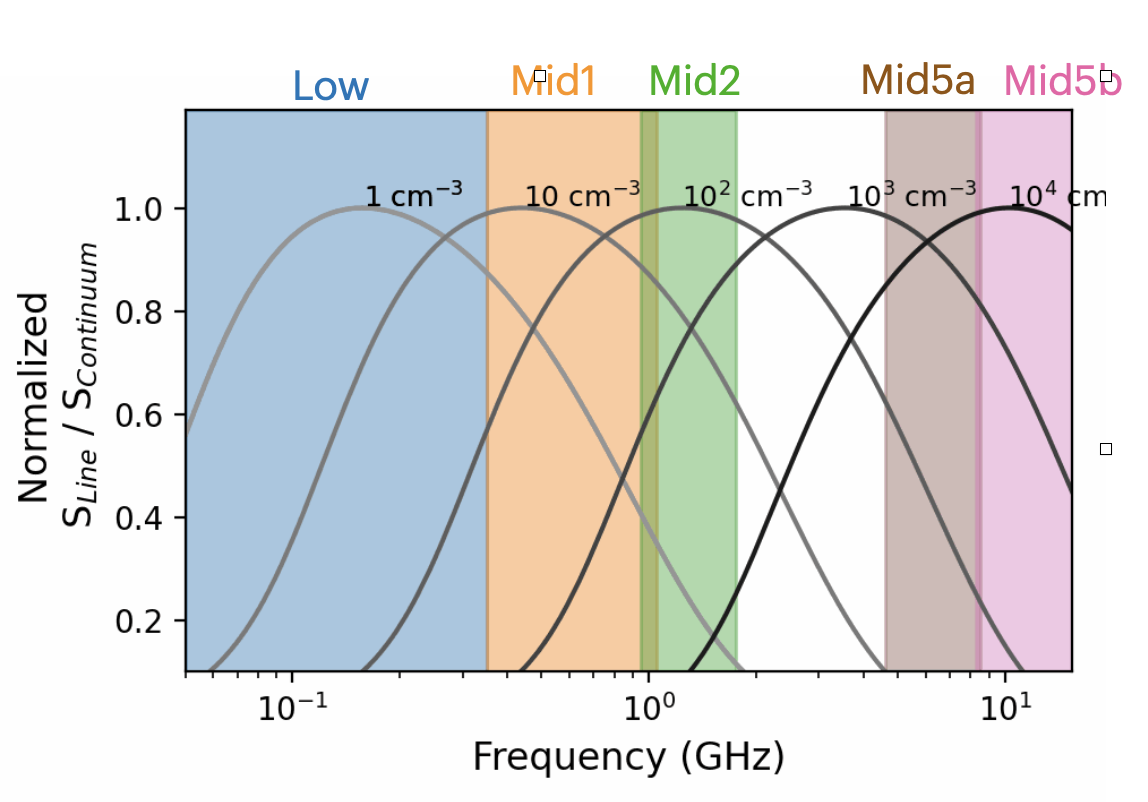}
    \includegraphics[width=0.48\linewidth]{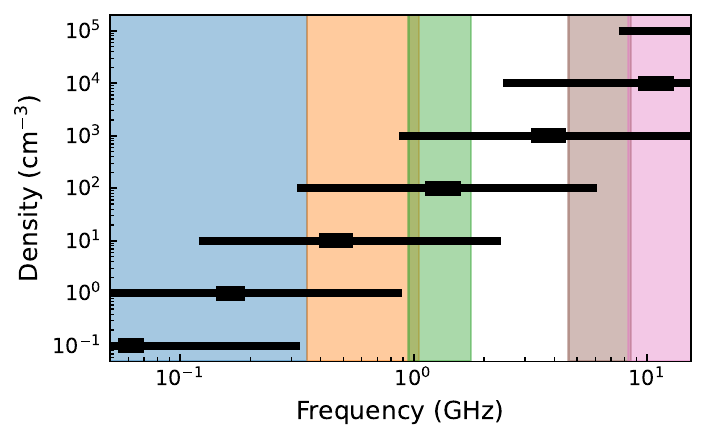}
    \caption{\textit{Left:} The normalized peak intensity of HRRLs shown for a range of densities (and fixed temperature). The frequency coverage of SKA AA* bands are indicated. While more than 10s of RRLs are covered in each band, the RRL intensity measured will depend on the characteristic density of the gas. \textit{Right:} For each order of magnitude density, the black line shows the frequencies covering the FWHM of the peak normalized peak intensity. Variation in the frequencies covered due to gas temperature (at a fixed density) is shown by a thicker black line; the effect of temperature on varying the frequency peaks of line intensities is minimal compared with density.}
    \label{fig:DIG_HRRLs}
\end{figure}

In Figure~\ref{fig:DIG_HRRLs} we show examples of the line-to-continuum ratio for HRRL for select densities of a gas slab with uniform density and a fixed temperature ($T_e = 8000$\,K). These plots show that the SLED of a single/characteristic component spans about a decade in frequency. For example, to sample with 4-5 frequency measurements gas described by a density of 100~\cmc{}, ideally one could use SKA AA* to measure the average RRL line intensity with two band measurements in SKA Mid Band 1, one average value in Band 2, and one average value in Band 5a.

In Figure~\ref{fig:DIG_HRRLs} we also plot the frequency at which each of the gas densities peaks; the total span of the bar indicates the frequencies which the line-to-continuum ratio is $\geq0.5$x the peak value. The span of the thicker inner bar indicates where the variation due to gas temperature for temperatures in the range 4000 -- 13\,000~K; density is most impactful to (and constrained by) the RRL line intensities; they are not strongly dependent upon temperature (of typical warm ionized gas temperatures).

RRLs always fall in the SKA bands, but the line emission would nominally probe different densities as function of $z$ for a given band. We list the nominal ionized gas electron density probed by HRRLs in each SKA AA* band for $z=0$ and $z=2$ sources in Table~\ref{tab:ska_hrrl_density}.

\begin{table}[]
    \centering
    \caption{Nominal (ionized) gas density probed in each SKA AA* band with HRRLs}
    \begin{tabular}{c|ccccc}
        \toprule
        & Low & Mid1 & Mid2 & Mid5a & Mid5b  \\
        \midrule
        $z=0$ & 1 \cmc{}   & 10 \cmc{}   & $10^2$ \cmc{} & $10^3$ \cmc{}  & $10^4$ \cmc{} \\
        $z=2$ & 10 \cmc{} & $10^2$ \cmc{} & $10^3$ \cmc{} & $10^4$ \cmc{} & $10^5$ \cmc{} \\
        \bottomrule
    \end{tabular}
    \label{tab:ska_hrrl_density}
\end{table}

\subsection{The environments and evolution of (compact) radio AGN}
\label{ssec:AGN}

Detections of low-frequency RRLs \textit{associated} with radio-bright continuum sources have still been limited to the nearby universe -- typically within a few Mpc. In absorption line studies, ``associated'' lines may be considered within 2000\,\kms{} \citep[e.g.,][]{Curran2021}. Detecting hydrogen and/or carbon RRLs associated with radio-bright continuum sources could be highly beneficial for investigating: narrow line regions, gas in circumnuclear disks of AGN, compact and peaked-spectrum AGN, jet interactions with the ISM, Lyman-$\alpha$ halos, ionized outflow material, and obscured quasars \citep[see SKA predictions by][]{Manti2016}. These investigations may also be highly complementary to full-polarization observations, which probe the rotation measure  of the ionized plasma ($RM \propto n_e \ell B_{||}$), whereas the RRLs probe the emission measure ($EM \propto n_e^2 \ell$).

Here we go into more detail about one of the most promising avenues, \textit{compact and peaked-spectrum AGN} \citep{ODea1998,ODea2021}. An important piece of evidence about the growth and evolution of super massive black holes (SMBH) comes from the radio emission that they (may or may not) emit. Furthermore, the radio emission from ``radio-loud'' AGN appears in various flavors. Some sources have extended lobes (where the lobe emission is bright with respect to emission from the narrow [inner] collimated jet), while others are more compact and may also exhibit a turnover in their SEDs. 
The compact sources out-number the extended sources, and there are more compact sources than would be expected if all compact sources evolved into extended sources. 

What dictates the lifetime of compact sources and/or hinders their evolution into large extended sources? Two leading ideas suggest this is regulated by (a) external mechanisms, such as the surrounding ISM which frustrate the growth of a (not powerful enough) compact source, vs (b) properties directly linked to the SMBH (feeding) such as the magnetic field and particle density \citep[e.g.,][]{Marscher1980}. These leading camps have largely stemmed from the possible mechanisms responsible for the SED turnovers in the `peaked-spectrum' compact sources -- (a) free-free absorption, vs (b) synchrotron self-absorption, respectively \citep{kellermann1966}. Because the two mechanisms have similar SED signatures, they have been difficult to distinguish observationally. HRRL detections from peaked-spectrum and compact sources will characterize the ionized gas properties (density, temp, pressure), which also give rise to free-free emission and absorption. HRRLs thus enable constraints on the 'young and frustrated' effect on the evolution of the AGN. 

Results from studies, albeit limited to small number statistics, suggest abundant HRRL detections. They suggest that a substantial number of peaked-spectrum sources are young, recently-triggered radio jets that are confined by interaction with dense gas in their central 1-2\,kpc of their host galaxy \citep[e.g.][]{ODea2021}. 
Furthermore, this makes them excellent candidates for CRRL detections from cold, molecular forming gas. PS objects often show 21\,cm HI absorption \citep[e.g.][]{pihlstrom2003,vermeulen2003,Yoon2025}.

\section{SKA Extragalactic RRL surveys: Observational Requirements}
\label{sec:obs}

We envision wide-band spectral-line surveys with the SKA that will target bright, $S \gtrsim 100$~mJy sources, including AGN and (nearby) star-forming galaxies. Ongoing surveys at 580--1670 MHz with SKA precursors are poised to lay the groundwork and bring about the most discovery potential with lower frequencies covered by the SKA, in the Low Band and Mid Band 1. We briefly sketch out observational requirements for what the first Square kilometer array Extragalactic RRL Surveys (SquERRLs) could look like with an hour per pointing integrations (see Table~\ref{tab:ska_survey}, and also Table~\ref{tab:rrl_nosources} for comparison with existing surveys; note, SKA's numbers are even more impressive if matching spatial resolutions but this also depends on the covering factor of the gas). Given that the native channel resolution across full bandwidth is suitable for extragalactic RRL observations, we highlight the high degree of commensal observing and scientific productivity that may be possible together with all-observable-sky continuum and/or full polarization surveys. For example, going from 1 hour integration per field to 8 hour integration per field\footnote{Here we point out that targeted observations, rather than survey} increases the number of detectable targets with the SKA-Low (SKA-Mid Band-1) from 167\,700 (102\,600) to 395\,600 (262\,200) as we can push down to continuum sources of $S \gtrsim 10$~mJy. 

With SKA-Low, 243 (169, 127) RRLs at $z= 0$ (2, 6) will be instantaneously covered. As the redshift of the emission region increases, transitions of different principal quantum numbers shift into the band; at $z=0$, $320\alpha$ is closest to the band midpoint (200 MHz), while for a $z=2$ source, $222\alpha$ is. At $z=0$, the RRLs are on average about 1.9 MHz apart (2800 \kms{}), but for $z=2$ they are about 2.7 MHz (4081 \kms{}); despite how numerous RRLs are in SKA's Low Band, RRLs are sufficiently far apart in frequency/velocity space that they will not overlap.
With SKA-Mid Band-1, 82 (56, 43) RRLs at $z= 0$ (2, 6) will be instantaneously covered. We referee the reader back to Section~\ref{sec:ska_impact} and Figure~\ref{fig:DIG_HRRLs} which describe how this additional line coverage is important for observing gas of different physical conditions and constraining its properties.

This science case is concerned about RFI over these frequencies, particularly in the unique coverage afforded by SKA, 350 - 580 MHz.

\begin{table}[htbp]
  \small
  \centering
  \caption{SquERRLs (Square kilometer array Extragalactic RRL Surveys) with AA4 \label{tab:ska_survey}}
    \begin{tabular}{lll}
    \toprule
    Parameter           & SKA-Low       & SKA-Mid Band-1\\
    \midrule
    Targets             & $S_{200\,\mathrm{MHz}} > 150$ mJy  & $S_{700\,\mathrm{MHz}} > 110$ mJy\\
    No. (detectable) Targets         & 167\,700  & 102\,600 \\
    Total Sky Area      & 34\,000 deg$^2$ & 34\,000 deg$^2$ \\
    Spatial Resolution  & 3.5$''$ -- 24.2$''$ & 0.51$''$ -- 1.54$''$ \\
    Time per pointing   & 1 hour & 1 hour \\
    $\nu_{\mathrm{obs}}$ & 50 -- 350 MHz & 350 -- 1050 MHz\\
    No. RRLs            & 243 ($z=0$), & 82 ($z=0$), \\
                        & 169 ($z=2$), & 56 ($z=2$),\\
                        & 127 ($z=6$)  & 43 ($z=6$)\\
    RRL $\alpha$ principal quantum no. & 508--266 ($z=0$), & 265--184 ($z=0$),\\
                              & 352--184 ($z=2$), & 183--128 ($z=2$), \\
                              & 265--139 ($z=6$) & 138--96 ($z=6$) \\
    $\Delta v_{\mathrm{chan}}$ & 4.6 -- 32.4 km s$^{-1}$ & 3.8 -- 11.5 km s$^{-1}$ \\
    $\sigma_{\mathrm{chan,\, requested}}$ & 1.06 mJy b$^{-1}$ & 0.53 mJy b$^{-1}$ \\
    $\sigma_{\mathrm{chan,\, stacked}}$ & 81 $\mu$Jy b$^{-1}$ & 69 $\mu$Jy b$^{-1}$ \\
    Commensal observing? & continuum; HI/OH abs; full pol & continuum; HI/OH abs; full pol\\
    \bottomrule
    \end{tabular}
\end{table}

\section{Conclusion}

The SKA provides a revolutionary new wealth of information by observing 1000s of accessible RRL transitions. We emphasize three areas of research in which detections of extragalactic low-frequency lines out to $z=6$ will have a high impact: ($\S$~\ref{ssec:HI-to-H2}) the \Hi{}-to-\Htwo{} transition across cosmic time, ($\S$~\ref{ssec:DIG}) dust-unobscured metallicity-independent (diffuse) ionized gas, and ($\S$~\ref{ssec:AGN}) the environments and evolution of (compact) AGN.  The SKA will be able to push extragalactic low-frequency RRL observations into a new frontier, by increasing the number of sources investigated from 1000s (upcoming in the next years) to 100\,000s. Wide-band large-area spectral-line surveys have great synergy between RRLs, \Hi{} absorption, and OH absorption to study galaxy evolution. The most discovery potential for RRLs with the SKA is expected to be (a) Low 50--350 MHz and (b) Mid Band1 350--1050 MHz. 

\bibliographystyle{abbrvnat-maxbibnames4}
\bibliography{chapter} 

\end{document}